\documentclass[aps,prl,showpacs,twocolumn,superscriptaddress]{revtex4}

\usepackage{graphicx}
\usepackage{epsfig}

\newcommand{\BEtwo}[0]{\mbox{B(E2; \ensuremath{0_{1}^{+}\rightarrow2_{1}^{+}})}}

\begin{document}

\title{Accuracy of \BEtwo{} transition rates from intermediate-energy Coulomb excitation experiments}

\author{J.M. Cook}
 \email[]{cook@nscl.msu.edu}
 \affiliation{National Superconducting Cyclotron Laboratory\\
   Michigan State University\\
   East Lansing, MI 48824}
 \affiliation{Department of Physics and Astronomy\\
   Michigan State University\\
   East Lansing, MI 48824}

\author{T. Glasmacher}
 \affiliation{National Superconducting Cyclotron Laboratory\\
   Michigan State University\\
   East Lansing, MI 48824}
 \affiliation{Department of Physics and Astronomy\\
   Michigan State University\\
   East Lansing, MI 48824}

\author{A. Gade}
 \affiliation{National Superconducting Cyclotron Laboratory\\
   Michigan State University\\
   East Lansing, MI 48824}

\date{\today}

\begin{abstract}
The method of intermediate-energy Coulomb excitation has been widely used to determine absolute \BEtwo~quadrupole excitation strengths in exotic nuclei with even numbers of protons and neutrons.  Transition rates measured with intermediate-energy Coulomb excitation are compared to their respective adopted values and for the example of \(^{26}\)Mg to the \BEtwo~values obtained with a variety of standard methods.  Intermediate-energy Coulomb excitation is found to have an accuracy comparable to those of long-established experimental techniques.
\end{abstract}


\maketitle

\section{\label{sec:intro}Introduction}
The electromagnetic transition matrix element \BEtwo{} between the ground state and the first \(2^+\) excited state of an even-even nucleus is a direct measure of the degree of quadrupole collectivity exhibited by the nucleus.  These reduced matrix elements are readily calculable in a variety of theoretical frameworks, such as the nuclear shell model~\cite{Bro01, Ots01} and different mean-field approaches (see e.g.\ \cite{Ben03, Lal99} and references within).  The experimental values of these reduced transition matrix elements have been compiled by Raman \cite{Raman}.  

Experimentally, transition matrix elements are accessible by measuring the lifetimes of excited states or the electromagnetic cross-sections to the excited states provided the excitation mechanism is understood. The latter approach has long been employed at energies below the Coulomb barrier~\cite{Alder1975} and was proposed 25 years ago for higher energies~\cite{Win79}.  Measurements of projectile Coulomb-excitation cross sections at beam energies well above the Coulomb barrier~\cite{RIKEN, Gla98} are ideal for rare-isotope experiments with low beam rates, which can be offset by reaction targets that are about 100--1000 times thicker than for below-barrier energies.  Post-target particle identification permits inverse-kinematic reconstruction of each projectile-target interaction.  Experiments at intermediate beam energies also allow for the unambiguous isotopic identification of incoming beam particles on an event-by-event basis, which is not generally possible at contemporary low-energy ISOL facilities.

Several reports on initial results from low-energy measurements on \(^{30}\)Mg~\cite{Sch05,Nie05a,Nie05b} questioned the accuracy of the intermediate-energy approach and speculated on possible error contributions.  The particular case of \(^{30}\)Mg now seems resolved in that the previously reported discrepancy~\cite{Ays05} has disappeared in the published low-energy result and is now in agreement with one intermediate-energy result~\cite{Pritychenko1999}, but not with another~\cite{Chi01}.  Responding to the general question raised, this paper examines the accuracy of the intermediate-energy Coulomb excitation method by comparing the ``test cases'' measured at intermediate beam energies at Michigan State University where an adopted reduced transition matrix element value based on four or more independent measurements with complementary techniques is available in the literature.  These test cases were measured over the past decade with the identical setups and during the same experiments used for measurements of unknown transition matrix elements.  While these test cases have been individually reported previously in peer-reviewed journals together with the respective new measurements, their collective comparison to adopted values here reaffirms intermediate-energy Coulomb excitation as an accurate method relative to other transition rate measurement techniques.

\section{\label{sec:measurements}Techniques for Measuring Transition Rates}
There are two general techniques for determining nuclear transition rates.  Lifetime measurements such as the Doppler shift attenuation method (DSAM) and the recoil distance Doppler shift (RDDS) method are based on the analysis of Doppler-shifted \(\gamma\)-ray peak shapes.  The lifetime~\(\tau\) of a state is related to the transition rate by
\begin{equation}
\label{eqn:lifetime}
\BEtwo{} \propto \frac{1}{E^{5}_{\gamma} \, \tau}
\end{equation}
for an E2 transition of energy \(E_{\gamma}\)~\cite{Bohr1998}.  In DSAM, nuclei are excited following fusion-evaporation reactions, Coulomb excitation, or inelastic scattering and, when the stopping power of the nuclei in the material is known, the Doppler shifts of the \( \gamma \) rays emitted by the recoiling reaction residues determine the points in time at which emission occured and hence the lifetime of the excited state (suitable for \(\tau<1\)~ps)~\cite{DSAM}.  The RDDS method similarly uses the Doppler shift of an excited, recoiling nucleus to determine the lifetime of the state.  A stopper is placed downstream of the target and the intensity ratio of $\gamma$ rays emitted in-flight and stopped for different target-stopper distances provides a measure of the lifetime in the range  \(10^{-9}\)~s~\( < \tau < 10^{-12}\)~s~\cite{RDDS}.

Nuclear resonance fluorescence (NRF), electron scattering, and Coulomb excitation determine transition rates through the measurement of cross sections.  In a typical NRF experiment, a continuous photon spectrum (bremsstrahlung) irradiates a target of stable nuclei.  The target nuclei are excited by the radiation and de-excitation $\gamma$~rays are subsequently emitted with an angular distribution depending on the transition.  The energy-integrated cross section of the scattered $\gamma$ rays is inversely proportional to the lifetime of the excited state~\cite{NRF}.  Electron scattering utilizes a simplified form of low-energy Coulomb excitation where the form factor in the Born approximation is related to the multipolarity of the transition.  The transition rate can be extracted from the value of the form factor~\cite{Rav}.  Due to the well-understood nature of the interaction and the ease of producing a large projectile flux, electron scattering is one of the most accurate methods of determining transition probabilities.  

\begin{figure}
  \epsfbox{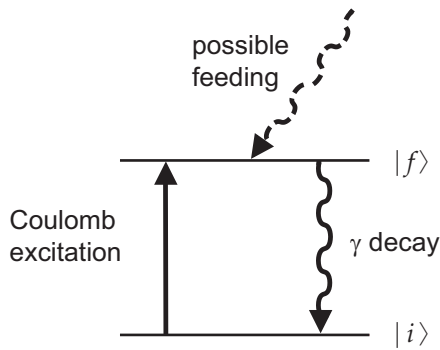}
\caption{\label{fig:leveldiagram}Schematic of Coulomb excitation of a nucleus from an initial state \( |i\rangle \) to a final bound state \( |f\rangle \) and the ensuing \(\gamma\) decay with a possible feeding transition from a higher state shown.}
\end{figure} 

In Coulomb excitation, the interaction of the electromagnetic fields of the target nuclei and projectile nuclei leads to excitations with subsequent \(\gamma\)-ray emissions.  The number of photons \(N_{\gamma, f \rightarrow i}\) observed in an inverse-kinematics Coulomb excitation experiment with \(\gamma\)-ray tagging is related to the excitation cross section by
\begin{equation}\label{eqn:ngamma}
\sigma_{i \rightarrow f}=\frac{N_{\gamma, f \rightarrow i}}{N_{T}N_{B}\epsilon}
\end{equation}
where $N_{T}$ is the number of target nuclei, $N_{B}$ is the number of beam nuclei, and \( \epsilon \) is the effeciency of the experimental setup.  $N_{B}$ can be determined prior to interaction with the target, and $N_{T}$ is given by the target thickness.  The efficiency accounts for the intrinsic and geometric efficiencies of all detector systems involved.  Equation~\ref{eqn:ngamma} assumes only one excited state; if more states than one are excited, possible feeding from higher excited states must be considered (see Figure~\ref{fig:leveldiagram}).  The excitation cross section can be related to the reduced transition probability through various approaches.  At beam energies below the Coulomb barrier of the projectile-target system, a Rutherford trajectory is assumed.  For intermediate-energy Coulomb excitation, we use the relativistic theory developed by Winther and Alder, which involves a semi-classical approach with first-order perturbation theory~\cite{Win79}.  Distorted-wave Born approximation calculations have also been used to determine transition rates from cross sections~\cite{RIKEN} and are in agreement with the excitation theory developed by Winther and Alder.

\section{\label{sec:coulex}Intermediate Energy Coulomb Excitation}

\begin{figure}
  \epsfbox{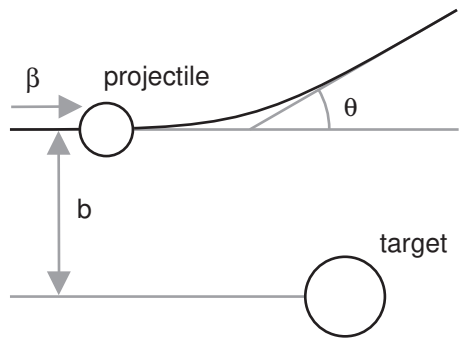}
\caption{\label{fig:angle}Schematic of a projectile nucleus scattering in the electromagnetic field of an infinitely heavy target nucleus.  For a fixed beam velocity \( \beta = v/c \), the scattering angle~\( \theta \) depends on the impact parameter b.  A maximum scattering angle is chosen in the experiment to restrict the minimum impact parameter.}
\end{figure}

The most important difference between low- and intermediate-energy Coulomb excitation is that nuclear interactions can occur above the Coulomb barrier.  However, the inclusion of nuclear contributions to the measurement of electromagnetic transition rate can be prevented in heavy-ion reactions by considering only those events scattered within a maximum scattering angle representing a ``safe'' minimum impact parameter~\(b_{min}\) (see Figure~\ref{fig:angle}).  The radius~\(R_{int}\) beyond which the Coulomb interaction dominates defines the minimum impact parameter to be allowed in the experiment.  Wilcke, et al.\ use elastic scattering data to predict \(R_{int}\) for interactions between various nuclei~\cite{Wilcke1980}.  For \(^{46}\)Ar it has been shown that varying \(b_{min}\) where \(b_{min} \geq R_{int}\) has little effect on the measured transition rate value~\cite{Gade03}.  In contrast, for light nuclei (approximately Z~\(<10\)) nuclear interactions may occur even for particles scattered at small angles and care must be taken to disentangle the nuclear and Coulomb contributions to the cross section~\cite{Gla01}.

The adiabatic cutoff of the Coulomb excitation process occurs at a maximum excitation energy
\begin{equation}
\label{eqn:adiabaticlimit}
E^{max}_{x} \approx \frac{\gamma \hbar c \beta}{b}
\end{equation}
where \( \beta = v/c \) and \( \gamma = \frac{1}{\sqrt{1- \beta^2}} \) are the velocity and Lorentz factor of the beam and \( b \) is the impact parameter.  Intermediate-energy beams can excite states at higher excitation energies compared to low-energy beams.  For example, \(^{26}\)Mg impinging on a \(^{209}\)Bi target with a beam velocity of \(\beta = 0.36\) has an adiabatic cutoff of \(E^{max}_{x} \approx 6\)~MeV~\cite{Jenthesis}.  However, the possibility of feeding from excitations to states above the first \(2^+\) state must be considered when calculating the excitation cross section~\cite{Gla98}.  Photons are used to identify the inelastic scattering process to bound excited states and hence target thickness is not constrained by the need to preserve momentum resolution to differentiate elastic and inelastic scattering.  Higher energy beams allow for the use of thicker targets, and the number of scattering centers can be increased by as much as a factor of 1000 over low-energy experiments, permitting an equivalent decrease in the number of required projectile nuclei.  In typical intermediate-energy Coulomb excitation experiments, 1~beam particle in \(10^3\)--\(10^4\) interacts with the target nuclei and multiple excitations are significant only to this small factor~\cite{Gla98}.  The wide range of scattering angles inherent in low-energy Coulomb scattering require large solid-angle detectors; a few degrees of acceptance suffices for intermediate-energy Coulomb excitation.  

At the NSCL, SeGA~\cite{Mue01}, an array of eighteen 32-fold segmented high-purity Ge $\gamma$-ray detectors, and APEX~\cite{Per03}, twenty-four position-sensitive NaI(Tl) crystals, are used for Coulomb-excitation measurements in conjunction with a phoswitch detector or the S800 spectrograph~\cite{Baz03} for event-by-event particle identification.  Similar setups are employed at GANIL~\cite{GANIL}, GSI~\cite{GSI/RISING}, and RIKEN~\cite{RIKEN}.  For a more detailed description of intermediate-energy Coulomb excitation, see~\cite{Win79, Gla98, Ber88}.

\section{\label{sec:data}Accuracy of Intermediate-energy Coulomb Excitation}
\begin{figure*}
  \epsfxsize 17.5cm
  \epsfbox{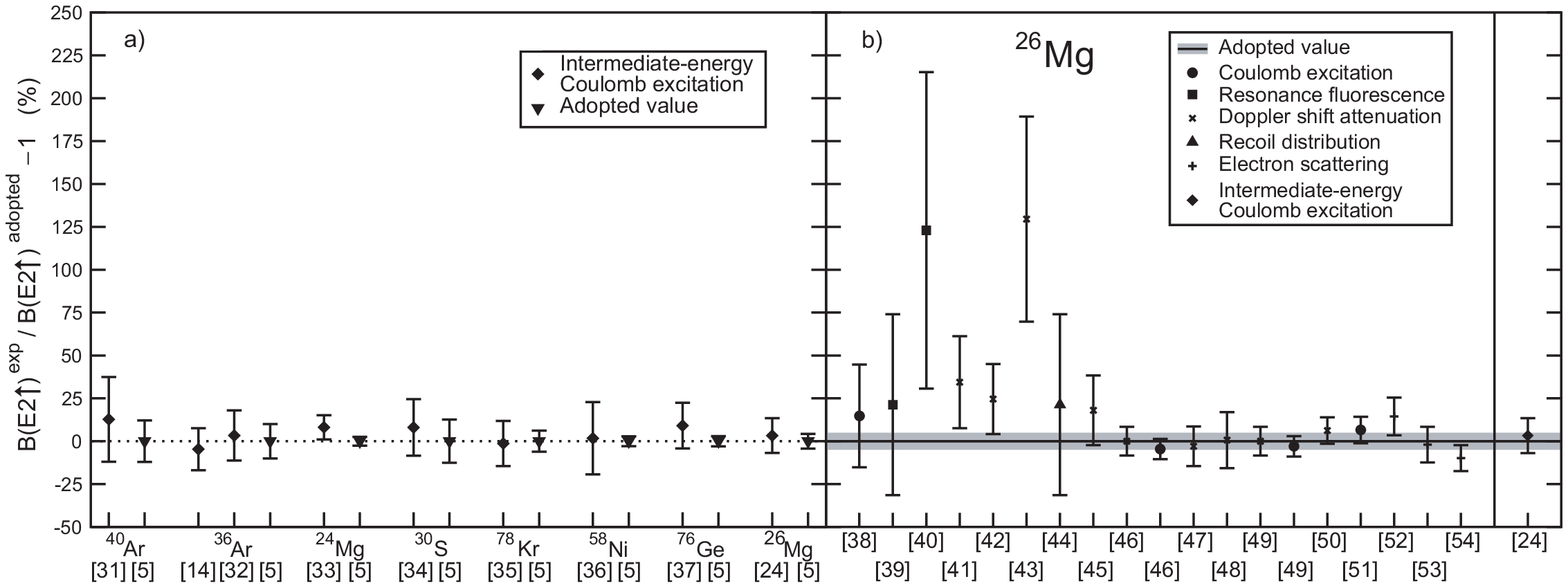}
\caption{\label{fig:composite}a) The percent differences between adopted and measured \BEtwo~transition rates for published test cases in intermediate-energy Coulomb excitation measurements. The average difference is 6\%~\cite{Ibbotson1998, Pritychenko1999, Cottle1999, Cottle2001, Cottle2002, Gade2005, Yurkewicz2004, Dinca2005, Jenthesis, Raman}.  b)~The percent differences between nineteen \BEtwo~transition rate measurements of $^{26}$Mg~\cite{And61, Ras51, Boo64, Hau68, Rob68, deK70, McD71, Dur72, Sch73, Ebe75, Wag1975, Sch1977, Dyb1981, Spe1982, Khv1970, Lee73, Lee74} and the adopted value~\cite{Raman} compared to an intermediate-energy Coulomb excitation measurement (right-most)~\cite{Jenthesis} of the same transition.  The 3\% difference of the intermediate-energy Coulomb excitation measurement compares favorably with the average absolute value of the difference of 23\% for the other measurements.}
\end{figure*}
The advantages of intermediate-energy Coulomb excitation are most pronounced when the method is applied to exotic nuclei with low production rates.  Under these circumstances, the statistical uncertainty dominates.  This difficulty is present irrespective of the method applied, and, therefore, only high-statistics intermediate-energy Coulomb excitation measurements will be considered in determining the method's accuracy.  A summary of intermediate-energy Coulomb excitation measurements of previously-published transition rates along with their respective adopted values can be found in Figure~\ref{fig:composite}a.  For these Coulomb excitation test cases, no feeding was observed.  The adopted values are those compiled by Raman~\cite{Raman} where four or more independent transition rate measurements using any of the above techniques have been made for each nucleus.  In the calculation of the adopted transition rate for \(^{40}\)Ar, one of eight experimental values was measured using intermediate-energy Coulomb excitation, and for \(^{36}\)Ar, two of eight.  The error bars on the adopted values represent the relative uncertainties.  The average difference from the adopted value is 6\% and only one data point exceeds 10\%. Note that all measurements are in agreement with their respective adopted values.

Figure~\ref{fig:composite}b shows the relative differences between measured \BEtwo~transition rates and the adopted value~\cite{Raman} for $^{26}$Mg.  The shaded area represents the uncertainty of the adopted value.  The measurements were made using low-energy (x,x'$\gamma$) Coulomb excitation, NRF, DSAM, RDDS, and electron scattering.  These traditional transition rate measurements have an average difference of 23\% from the adopted value for $^{26}$Mg.  The right-most data point, which deviates from the adopted value by 3\%, was measured by Church et al.\ \cite{Jenthesis} using intermediate-energy Coulomb excitation at a beam energy of 66.8~MeV/nucleon.
This specific measurement illustrates the more general point made in Figure~\ref{fig:composite}a that intermediate-energy Coulomb excitation measurements that are not limited by statistics can readily measure transition rates with an accuracy of about 5\% to a precision of about 10\%.

\section{\label{sec:conclusion}Conclusion}

The extention of the Coulomb excitation method to projectiles at intermediate beam energies allows for the measurement of transition rates in nuclei far from stability.  Thick targets allow for experiments on isotopes with low projectile fragmentation production rates.  Intermediate-energy Coulomb excitation has been shown to produce results within error of the adopted values for transitions measured with long-established experimental techniques.  Additionally, the accuracy of the \(^{26}\)Mg transition rate measurement from intermediate-energy Coulomb excitation exceeds the average accuracy of the other measurements.

\begin{acknowledgments}
This work is supported by the National Science Foundation through grants PHY-0110253.
\end{acknowledgments}

\end{document}